\documentclass[aps,superscriptaddress,twocolumn]{revtex4-2}
\usepackage[utf8]{inputenc}
\usepackage{amsthm,amssymb,amsmath}
\usepackage{graphicx,xcolor}
\usepackage[colorlinks=true,linkcolor=blue,citecolor=blue,urlcolor=blue,linktocpage=true]{hyperref}

\usepackage{microtype}
 \usepackage{siunitx}
\usepackage{float}
\usepackage{booktabs}

\usepackage{mathtools}

\begin{document}
\title{Purified pseudomode model for nonlinear system-bath interactions}

\author{Cheng Zhang}
\affiliation{Research Center for Quantum Physics and Technologies, Inner Mongolia University, Hohhot 010021, China}
\affiliation{School of Physical Science and Technology, Inner Mongolia University, Hohhot 010021, China}

\author{Neill Lambert}
\email{nwlambert@gmail.com}
\affiliation{RIKEN Center for Quantum Computing (RQC), Wakoshi, Saitama 351-0198, Japan}

\author{Xin-Qi Li}
\affiliation{Research Center for Quantum Physics and Technologies, Inner Mongolia University, Hohhot 010021, China}
\affiliation{School of Physical Science and Technology, Inner Mongolia University, Hohhot 010021, China}

\author{Mauro Cirio}
\email{cirio.mauro@gmail.com}
\affiliation{Graduate School of China Academy of Engineering Physics, Haidian District, Beijing, 100193, China}

\author{Pengfei Liang}
\email{pfliang@imu.edu.cn}
\affiliation{Research Center for Quantum Physics and Technologies, Inner Mongolia University, Hohhot 010021, China}
\affiliation{School of Physical Science and Technology, Inner Mongolia University, Hohhot 010021, China}

\date{\today}
\begin{abstract}
The theory of purified pseudomodes [arXiv:2412.04264 (2024)] was recently developed to provide a numerical tool for the analysis of the properties of a quantum system and the environment it couples to via linear system-bath interactions. Here we extend this theory to allow for the description of general nonlinear system-bath interactions. We demonstrate the validity of our method by considering the spontaneous decay of a two-level atom placed inside a single-mode lossy cavity and furthermore, its potential application to nanophotonics by calculating the resonance fluorescence spectrum of a quantum dot in the presence of a phonon environment. Our method provides a useful tool for the study of phonon-assisted emission in quantum dots and holds the promise for broad applications in fields like quantum biology, nonlinear phononics, and nanophotonics. 
\end{abstract}

\maketitle

\section{introduction}
The manipulation and synthesis of photon- and phonon-matter interactions have achieved considerable progresses in past decades, making some challenging regimes experimentally accessible. Representative examples include, e.g., ultra- and deep-strong light-matter interactions in cavity and circuit quantum electrodynamics (QED)~\cite{RevModPhys.91.025005,FriskKockum2019,QIN20241}, time retardation effects caused by light propagation in waveguide QED~\cite{Giuseppe,PhysRevLett.123.013601,PhysRevA.104.053701} and giant atoms~\cite{PhysRevA.90.013837,PhysRevLett.120.140404,Andersson2019,GuoOBS}, nonlinear optomechanical interactions induced by radiation pressure~\cite{PhysRevA.95.053861,Leijssen2017,PhysRevLett.131.053601,Burgwal2023}, and nonlinear phononics~\cite{WeiSS:21,Henstridge2022,Hackett2024}, to name a few. A faithful and accurate characterization of the properties of physical systems in these regimes requires specific analytical and numerical methods. 

The common starting point of devising such methods is often a microscopic description of the environment and its interaction with the system of interest \cite{Petruccione,Gardiner}. 
A variety of perturbative or non-perturbative treatments can be derived depending on the parameter regime of interest. For example, perturbation theory can lead to an effective description of the system in terms of quantum master equations~\cite{Petruccione,Gardiner,Lindblad1976,GORINI1978149,PhysRevLett.132.170402}.
On the other hand, to model non-pertubative regimes, more advanced techniques has to be used such as the discretization of the environmental continuum ~\cite{PhysRevB.92.155126,10.1063/1.5135363,Shuai_review} or the Feymann-Vernon influence functional~\cite{Strathearn_2017,Strathearn2018,PhysRevLett.123.240602,PhysRevLett.126.200401,PhysRevLett.129.173001,PRXQuantum.3.010321,PhysRevLett.132.200403,PhysRevX.14.011010} for tensor-network simulations, and the analysis of equivalent models defined by 
the reaction coordinate mapping~\cite{Garg,Martinazzo,Woods,iles2014environmental,PhysRevB.97.205405,Melina}, the chain mapping~\cite{Chin_2010,PhysRevLett.105.050404,PhysRevLett.123.090402,PhysRevB.101.155134} and the polaron transformation~\cite{Holstein1,Holstein2,Jackson,Silbey,Silbey2,Jang,Jang2}. In alternative, it is also possible construct effective models involving auxiliary degrees of freedom, e.g., the auxiliary density matrices in the hierarchical equations of motion (HEOM)~\cite{Tanimura_3,Tanimura_1,Ishizaki_1,Tanimura_2,Ishizaki_2,PhysRevLett.109.266403,PhysRevA.85.062323,Tanimura_2014,Tanimura_2020,FreePoles}, dissipatons~\cite{Yan_1,Yan_2,10.1063/5.0123999,10.1063/5.0151239}, and pseudomodes~\cite{garraway1997,Dorda,Mascherpa,Tamascelli,Lambert2019,PhysRevA.101.052108,PhysRevResearch.2.043058,PhysRevLett.126.093601,PRXQuantum.4.030316,PhysRevLett.132.106902,PhysRevB.110.195148,PhysRevA.110.022221,PhysRevResearch.6.033237,albarelli2024}.

In a previous work~\cite{liang2024purifiedinputoutputpseudomodemodel}, based on the pseudomode theory~\cite{garraway1997,Dorda,Mascherpa,Tamascelli,Lambert2019,PhysRevA.101.052108,PhysRevResearch.2.043058,PhysRevLett.126.093601,PRXQuantum.4.030316,PhysRevLett.132.106902,PhysRevB.110.195148,PhysRevA.110.022221,PhysRevResearch.6.033237,albarelli2024}, some of us introduced a new class of ancillary damping modes and detailed a rigorous scheme to construct effective models for the numerical simulation of the reduced dynamics of a system and the input-output properties of its environment. These ancillary modes were named ``purified pseudomodes" because their states remain pure during the dynamics, 
in stark contrast to pseudomodes which require a mixed-state description. This method is numerically exact and valid for thermal bosonic environments and for generic linear system-bath interactions, and only requires the spectral decompositions of the bath two-time correlation matrix. 

Here, we derive an extended purified pseudomode model for the simulation of open quantum systems in the presence of general nonlinear system-bath interactions, where the bath coupling operator is an analytical function of a linear coupling operator $X$, see the definition in Eq.~(\ref{eq:HintsQ}). This topic remains largely unexplored in the literature, despite the notable results in Refs.~\cite{10.1063/1674-0068/30/cjcp1706123,10.1063/1.4991779} where the specific case of a system-bath interaction written as the sum of a linear term and a quadratic correction was considered in the context of the HEOM. 
The analysis of more general nonlinear system-bath interactions, including the specific ones employed in  Refs.~\cite{10.1063/1674-0068/30/cjcp1706123,10.1063/1.4991779}, is more challenging than the linear case because of the 
additional terms emerging in the  Wick factorizations of multi-time bath correlation functions which depend on non-trivial equal-time  contractions. In this article, we show that the technical difficulties involved in re-summing all these extra Wick contractions due to nonlinearities are bypassed using the pseudomode theory. In fact, by its very definition, this method offers a natural one-to-one correspondence of multi-time correlation functions between the original bath and the corresponding pseudomode model, allowing the latter to encode any nonlinear statistics of the original bath by mimicking the statistical properties of the linear operator $X$. Despite its generality, the resulting algorithm is easy implementable for the numerical simulation of open quantum systems with non-linear couplings. We numerically illustrate the validity of this method to model the spontaneous decay of a two-level atom strongly coupled to a single-mode lossy cavity. We introduce further potential applications of our method to more complex cases by studying the effects of nonlinearly-coupled phonons on the fluorescence spectrum of a two-level quantum dot. 

The paper is organized as follows. In Sec.~\ref{sec:model}, we describe the microscopic model of an open quantum system in which a system couples to a bosonic environment with nonlinear interactions, and show that extra bath statistics is required to reproduce the reduced state of the system. In Sec.~\ref{sec:PPMconstruction}, we present the construction of an explicit purified pseudomode model allowing for the simulation of the reduced system dynamics. In Sec.~\ref{sec:numerics}, we numerically demonstrate the validity and applications of our method with two examples. A summary of  the main results and an outlook are given in Sec.~\ref{sec:summary}.

\section{model and reduced system dynamics}\label{sec:model}

We start by defining the general model of an open quantum system consisting of a system $S$ and a bosonic environment $B$, described by the Hamiltonian
\begin{equation}
H = H_S + H_B + H_{\text{int}}, 
\end{equation}
where $H_S$ is the system Hamiltonian and $H_B = \sum_k \omega_k b_k^\dagger b_k$ is the bath Hamiltonian with $b_k$ ($b_k^\dagger$) denoting the annihilation (creation) operator of the bath harmonic mode having frequency $\omega_k$, satisfying $[b_k,b_{k'}^\dagger]=\delta_{kk'}$. We consider the following general form for the 
system-bath interaction
\begin{equation}\label{eq:HintsQ}
H_{\text{int}} = sQ(X) = s\sum_{n=1}^\infty \alpha_n X^n, 
\end{equation}
where $s$ and $Q$ are the system and bath  coupling operators respectively, and $X=\sum_k(g_kb_k^\dagger + g_k^*b_k)$ is a linear bath coupling operator with $g_k$ the coupling amplitudes. Here we have assumed that the bath coupling operator $Q$ is an analytical function of $X$, in other words, $Q$ can be expanded as a power series of $X$ with real-valued coefficients $\alpha_n\in\mathbb{R}$, so that $Q$ is Hermitian. In particular, the case $\alpha_1=1$ and $\alpha_{n\ge2}=0$ corresponds to the linear system-bath interaction considered in most literature. In Refs.~\cite{10.1063/1674-0068/30/cjcp1706123,10.1063/1.4991779}, Xu and collaborators analyzed the case $\alpha_{n\ge3}=0$ in the context of the HEOM. 

We are interested in the reduced dynamics of the system $S$, encoded in the reduced density operator $\rho_S(t) \equiv \text{Tr}_B \rho(t)$, 
which is obtained by tracing over the environmental degrees of freedom in the full density matrix $\rho(t)$. The evolution of $\rho(t)$ obeys the von-Neumann equation $d\rho(t)/dt = -i[H,\rho(t)]$ in the Schr\"odinger picture. We also assume a factorizing initial state $\rho(0) = \rho_S(0)\otimes\rho_{\text{th}}$, where $\rho_{\text{th}} = \exp(-\beta H_B)/\text{Tr}_B\exp(-\beta H_B)$ denotes the bath thermal state at inverse temperature $\beta$. 

In the interaction picture with respect to the free Hamiltonian $H_0 = H_S + H_B$, the reduced system dynamics is described by
\begin{equation}\label{eq:rhoSeomIP}
\frac{d\rho_S^I(t)}{dt} = -i\text{Tr}_B[H_{\text{int}}^I(t), \rho_S^I(t)], 
\end{equation}
where $\rho_S^I(t) = e^{iH_0t}\rho_S(t)e^{-iH_0t}$ and $H_{\text{int}}^I(t) = e^{iH_0t}H_{\text{int}}e^{-iH_0t}$ denote the corresponding reduced density operator and system-bath interaction Hamiltonian in the interaction picture, respectively. The formal solution of Eq.~(\ref{eq:rhoSeomIP}) can be written \cite{PRXQuantum.4.030316} as the Dyson series
\begin{equation}\label{eq:rhoSformalseries}
\rho_S^I(t) = \sum_{m=0}^\infty\frac{(-i)^m}{m!}\rho_m^I(t), 
\end{equation}
where the $m$th-order contribution is
\begin{equation}\label{eq:rhon}
\begin{array}{lll}
\rho_m^I(t) &\displaystyle = \int_0^tdt_m\cdots\int_0^tdt_1 \sum_{i_1,\cdots,i_m\in\{l,r\}}\langle\mathcal{Q}_{t_m}^{i_m}\cdots\mathcal{Q}_{t_1}^{i_1}\rangle_{\text{th}} \\
&\displaystyle~~~\times  \mathcal{T}_S\mathcal{S}_{t_n}^{i_n}\cdots\mathcal{S}_{t_1}^{i_1}\rho_S(0), 
\end{array}
\end{equation}
with $\langle\mathcal{Q}_{t_m}^{i_m}\cdots\mathcal{Q}_{t_1}^{i_1}\rangle_{\text{th}}\equiv\text{Tr}_B\Big[\mathcal{T}_B\mathcal{Q}_{t_m}^{i_m}\cdots\mathcal{Q}_{t_1}^{i_1}\rho_{\text{th}}\Big]$ denoting the multi-time bath correlation functions and the indexes $i_1,\cdots,i_m\in\{l,r\}$ labeling the left- or right-action of the superoperators $\mathcal{S}_{i_j}^{t_j}$ and $\mathcal{Q}_{t_j}^{i_j}$. In other words, they are defined as $\mathcal{S}_{t_j}^l[\cdot] = s(t_j)[\cdot]$, $\mathcal{S}_{t_j}^r[\cdot] = -[\cdot]s(t_j)$, $\mathcal{Q}_{t_j}^l[\cdot] = Q(t_j)[\cdot]$, and $\mathcal{Q}_{t_j}^r[\cdot] = [\cdot]Q(t_j)$ where $s(t_j)=\exp(iH_St_j)s\exp(-iH_St_j)$ and $Q(t_j)=\exp(iH_Bt_j)Q\exp(-iH_Bt_j)$ correspond to the operators $s$ and $Q$ in the interaction picture, respectively. Note that, the minus sign in the definition of $\mathcal{S}_{t_j}^r$ is necessary to recover the commutators involving $H_{\text{int}}^I(t)$ in the Dyson series in Eq.~(\ref{eq:rhoSformalseries}). The time-ordering operators $\mathcal{T}_{S}$ and $\mathcal{T}_{B}$ respectively ensure that  the superoperators $\mathcal{S}_{i_j}^{t_j}$ and $\mathcal{Q}_{t_j}^{i_j}$ are in chronological order. 

In the case of linear system-bath interaction $Q=X$, the superoperator version of Wick's theorem proved in Ref.~\cite{PRXQuantum.4.030316} allows to conclude that each multi-time correlation function $\langle\mathcal{Q}_{t_m}^{i_m}\cdots\mathcal{Q}_{t_1}^{i_1}\rangle_{\text{th}}$ appearing in Eq.~(\ref{eq:rhon}) can be written as  a sum of total contraction of the linear operator $X$. More precisely, each multi-time correlation function $\langle\mathcal{Q}_{t_n}^{i_n}\cdots\mathcal{Q}_{t_1}^{i_1}\rangle_{\text{th}}$ can be reduced to a product of two-time correlators which are related to the bath two-time correlation function
\begin{equation}\label{eq:CXX}
\begin{array}{lll}
\displaystyle C(t) &\displaystyle \equiv \text{Tr}_B\left[ X(t_2)X(t_1)\rho_{\text{th}}\right]  \\
&\displaystyle = \int_0^\infty d\omega \frac{J(\omega)}{\pi} \left[\coth(\beta\omega/2)\cos(\omega t) - i\sin(\omega t) \right], 
\end{array}
\end{equation}
where $t=t_2-t_1$ denote the time difference and $J(\omega)\equiv\pi\sum_k\lvert g_k\rvert^2\delta(\omega-\omega_k)$ is the spectral density. The time-translational invariance of $C(t)$ is a consequence of the stationarity of the bath initial state $\rho_{\text{th}}$ under the bath free dynamics, manifested in the commutation relation $[H_B,\rho_{\text{th}}]=0$. In addition, $C(t)$ satisfies the time-reversal property $C(t)=C^*(-t)$ with the symbol $*$ representing the complex conjugation. Ultimately, the formal series in Eq.~(\ref{eq:rhoSformalseries}) can be re-summed to yield a compact expression $\rho_S^I(t) = \exp(-\mathcal{F}(t,s,C(t))\rho_S(0)$ where the influence superoperator $\mathcal{F}(t,s,C(t))$ is explicitly written as~\cite{PRXQuantum.4.030316}
\begin{equation}\label{eq:Ft}
\begin{array}{lll}
\mathcal{F}[\boldsymbol\cdot] &\equiv&\displaystyle 
\int_0^tdt_2\int_0^{t_2}dt_1 \Big\{ C(t_1-t_2)[\boldsymbol\cdot s(t_1),s(t_2)] \\
&&\displaystyle\phantom{\int_0^tdt_2\int_0^{t_2}dt_1 \Big\{} - C(t_2-t_1)[s(t_1)\boldsymbol\cdot,s(t_2)] \Big\}. 
\end{array}
\end{equation}
This compact expression, or equivalently its path integral representation, constitutes the starting point of several nonperturbative methods~\cite{Strathearn_2017,Strathearn2018,PhysRevLett.123.240602,PhysRevLett.129.173001,PhysRevLett.126.200401,PRXQuantum.3.010321,PhysRevLett.132.200403,PhysRevX.14.011010,Tanimura_1,Tanimura_2,Tanimura_3,Tanimura_2020,Tanimura_2014,Ishizaki_1,Ishizaki_2,FreePoles,PhysRevLett.109.266403,PhysRevA.85.062323,garraway1997,Dorda,Mascherpa,Tamascelli,Lambert2019,PhysRevA.101.052108,PhysRevResearch.2.043058,PhysRevLett.126.093601,PRXQuantum.4.030316,PhysRevLett.132.106902,PhysRevB.110.195148,PhysRevA.110.022221,PhysRevResearch.6.033237,albarelli2024} developed for linear system-bath interaction. 

While no compact expression can be found for $\rho_S^I(t)$ in the general case of nonlinear system-bath interactions considered in Eq.~(\ref{eq:HintsQ}), Wick's theorem~\cite{PRXQuantum.4.030316} still allows to conclude that each multi-time correlation function $\langle\mathcal{Q}_{t_n}^{i_n}\cdots\mathcal{Q}_{t_1}^{i_1}\rangle_{\text{th}}$ in Eq.~(\ref{eq:rhon}) can be written as a sum over the product of contractions involving the linear operator $X$, i.~e., in terms of the bath correlation function $C(t)$. In the nonlinear case, the resulting factorizations of 
$\langle\mathcal{Q}_{t_n}^{i_n}\cdots\mathcal{Q}_{t_1}^{i_1}\rangle_{\text{th}}$ also acquire explicit dependencies on the second moment of $X$, i.e. 
\begin{equation}\label{eq:X2}
\sigma^2 \equiv \text{Tr}_B\left[ X(t_1)X(t_1)\rho_{\text{th}}\right] = C(0).
\end{equation}
As an example, in the quadratic case $Q=X^2$, the factorization of $\langle\mathcal{Q}_{t_2}^{i_2}\mathcal{Q}_{t_1}^{i_1}\rangle_\text{th}$ reads 
\begin{equation}
\begin{array}{lll}
\langle\mathcal{Q}_{t_2}^{i_2}\mathcal{Q}_{t_1}^{i_1}\rangle_\text{th} &\displaystyle= \langle \mathcal{X}_{t_2}^{i_2}\mathcal{X}_{t_2}^{i_2}\rangle_\text{th} \langle \mathcal{X}_{t_1}^{i_1}\mathcal{X}_{t_1}^{i_1}\rangle_\text{th} \\
&\displaystyle~~~ + 2\langle \mathcal{X}_{t_2}^{i_2}\mathcal{X}_{t_1}^{i_1}\rangle_\text{th} \langle \mathcal{X}_{t_2}^{i_2}\mathcal{X}_{t_1}^{i_1}\rangle_\text{th}, 
\end{array}
\end{equation}
where $\mathcal{X}_{t_n}^l[\cdot] = X({t_n})[\cdot]$, $\mathcal{X}_{t_n}^r[\cdot] = [\cdot]X({t_n})$ with $n=1,2$. The moment $\sigma^2$ and the two-time correlation function $C(t)$ appear in the first and second lines, respectively. 
This explicit dependence on the moment $\sigma^2$ in the Wick factorizations of $\langle\mathcal{Q}_{t_n}^{i_n}\cdots\mathcal{Q}_{t_1}^{i_1}\rangle_{\text{th}}$, has critical and profound consequences on the construction of the purified pseudomode model, constituting a major challenge which we overcome in the next section.

\section{Purified pseudomode model for nonlinear system-bath interactions}\label{sec:PPMconstruction}

In this section, we explicitly construct a purified pseudomode model that allows to reproduce the reduced system dynamics of the nonlinear open-system model described in Sec.~\ref{sec:model}. To this end, we first present a brief summary of the conventional pseudomode theory in Sec.~\ref{sec:pmtheory}. We then establish an explicit pseudomode model that reproduces both the two-time correlation in Eq.~(\ref{eq:CXX}) and the second moment in Eq.~(\ref{eq:X2}) in Sec.~\ref{sec:PMmodel}, based on spectral decompositions of the two-time correlation in Eq.~(\ref{eq:CXX}). In Sec.~\ref{sec:purification}, we generalize the purification technique presented in our previous work~\cite{liang2024purifiedinputoutputpseudomodemodel} to reach a purified pseudomode model for the special case $Q=X^n$ and furthermore, for the general nonlinear system-bath interaction~(\ref{eq:HintsQ}).

\subsection{The pseudomode theory}\label{sec:pmtheory}

In the conventional pseudomode theory~\cite{garraway1997,Tamascelli,Lambert2019}, where often only linear system-bath interaction $Q=X$ is considered, the original continuum  environment is modeled as an effective, discrete one so that they yield identical reduced dynamics of the system $S$. The effective environment consists of a number of discrete ancillary bosonic modes, each subject to an additional Markovian dissipation channel, so that the state $\rho_\text{pm}(t)$ describing the original system $S$ and the effective environment obeys the following Lindblad master equation in the Schr\"odinger picture 
\begin{equation}\label{eq:PMgeneralmodel}
\begin{array}{lll}
\displaystyle \frac{d\rho_\text{pm}}{dt} &\displaystyle = -i[H_\text{S} + \sum_\alpha\Omega_\alpha d_\alpha^\dagger d_\alpha + sQ_\text{pm},\rho_\text{pm}] \\
&\displaystyle~~~ + \sum_{\alpha}\Gamma_{\alpha}(2d_\alpha\rho_\text{pm}d_\alpha^\dagger - d_\alpha^\dagger d_\alpha\rho_\text{pm} - \rho_\text{pm} d_\alpha^\dagger d_\alpha), 
\end{array}
\end{equation}
where the coupling operator $Q_\text{pm}= X_\text{pm}$ is related to the linear coupling operator 
\begin{equation}\label{eq:XPM}
X_\text{pm}\equiv \sum_{\alpha}X_{\text{pm},\alpha} = \sum_\alpha\lambda_\alpha(d_\alpha+d_\alpha^\dagger), 
\end{equation}
and $d_\alpha$ ($d_\alpha^\dagger$) denote the annihilation (creation) operator of the $\alpha$th pseudomode with frequency $\Omega_\alpha$ and decay rate $\Gamma_\alpha$. In Eq.~(\ref{eq:PMgeneralmodel}), we have  used a simplified version of the general pseudomode model~\cite{Tamascelli,Lambert2019,PRXQuantum.4.030316}, namely, each pseudomode is in contact with its own background environment held at zero temperature, resulting in a single Lindblad dissipator with decay rate $\Gamma_\alpha$ in the second line of Eq.~(\ref{eq:PMgeneralmodel}). For consistency, this also implies that the initial state of the pseudomode model in Eq.~(\ref{eq:PMgeneralmodel}) should be $\rho_\text{pm}(0)=\rho_S\otimes\rho_\text{vac}$ with $\rho_\text{vac}=\otimes_\alpha|0_\alpha\rangle\langle0_\alpha|$ the vacuum state of all pseudomodes. Note that this zero-temperature assumption, alongside the independent pseudomode assumption underlying Eq.~(\ref{eq:PMgeneralmodel}), can both be relaxed to further enlarge predictivity of the pseudomode model~\cite{Lambert2019,PhysRevA.101.052108,PRXQuantum.4.030316,PhysRevLett.126.093601,PhysRevLett.132.106902}. For example, all model parameters can be analytically continued to the complex plane such that they become complex-valued, i.e., $\Omega_\alpha,\Gamma_\alpha,\lambda_\alpha\in\mathbb{C}$~\cite{Lambert2019,PRXQuantum.4.030316,PhysRevResearch.6.033237}. On the other hand, mode-mode interactions can also be introduced to form a pseudomode network~\cite{PhysRevA.101.052108,PhysRevLett.126.093601,PhysRevLett.132.106902}. 

The equivalence between the original environment $B$ and the effective one made of pseudomodes is guaranteed if the latter can reproduce (at least approximately) the original bath correlation function $C(t)$~(\ref{eq:CXX}), i.e., 
\begin{equation}\label{eq:eqvcondition}
C(t) \cong C_\text{pm}(t), 
\end{equation}
where $C_\text{pm}(t) \equiv \text{Tr}_\text{pm}[X_\text{pm}(t)X_\text{pm}(0)\rho_\text{vac}]$ can be solved analytically as~\cite{PRXQuantum.4.030316}
\begin{equation}\label{eq:CPM}
C_\text{pm}(t) = \sum_\alpha\lambda_\alpha^2 e^{-i\Omega_\alpha t - \Gamma_\alpha\lvert t\rvert}. 
\end{equation}
In other words, for linear system-bath interaction, this equivalence condition ensures that the reduced density matrix $\rho_S(t)$ for the open-system model in Sec.~\ref{sec:model} can also be obtained by tracing out in $\rho_\text{pm}(t)$ all pseudomode degrees of freedom, i.e., $\rho_S(t) = \text{Tr}_\text{pm}\rho_\text{pm}(t)$.

The above consideration can be easily generalized to nonlinear system-bath interactions~(\ref{eq:HintsQ}). This is achieved by simply substituting 
\begin{equation}\label{eq:QPM}
Q_\text{pm}=\sum_{n=1}^\infty \alpha_n X_\text{pm}^n
\end{equation}
in Eq.~(\ref{eq:PMgeneralmodel}).
The formal correspondence between $Q_\text{pm}$ and the original bath coupling operator $Q(X)$ in Eq.~(\ref{eq:HintsQ}) has to be such to establish the correspondence  $\langle\mathcal{Q}_{t_m}^{i_m}\cdots\mathcal{Q}_{t_1}^{i_1}\rangle_{\text{th}} = \langle\mathcal{Q}_{\text{pm},t_m}^{i_m}\cdots\mathcal{Q}_{\text{pm},t_1}^{i_1}\rangle_{\text{th}}$ between the multi-time correlation functions of the full and effective models (here, the superoperators $\mathcal{Q}_{\text{pm},t_m}^{i_m}$ are defined analogously as $\mathcal{Q}_{t_m}^{i_m}$). Using Wick's theorem,  we can then directly conclude that all these equivalences hold when Eq.~(\ref{eq:eqvcondition}) is imposed. 

It is important to note that the equivalence condition in Eq.~(\ref{eq:eqvcondition}) is not required to be pointwise, i.e., to be valid for all $t\in\mathbb{R}$. More precisely, it can be allowed to be weakly violated, i.e., such that $C(t)\neq C_\text{pm}(t)$ over a countable set of times $\mathcal{S}= \{t\in\mathbb{R}|C(t)\neq C_\text{pm}(t)\}$. This freedom arises from the property of Riemann integrals and the fact that $C(t)$ always appears as an integrand in Eq.~(\ref{eq:rhon}). However, we emphasize that the explicit emergence of the second moment in Eq.~(\ref{eq:X2}) in the Wick factorizations of the multi-time correlation function $\langle\mathcal{Q}_{t_m}^{i_m}\cdots\mathcal{Q}_{t_1}^{i_1}\rangle_{\text{th}}$ for the nonlinear interaction~(\ref{eq:HintsQ}) requires a precise approximation of the correlation at $t=0$, i.e., $(t=0)\notin\mathcal{S}$, in stark constrast to the linear case. In Sec.~\ref{sec:PMmodel} we show how the freedom described here facilitates the imposition of this constraint at $t=0$ by adding an extra zero-frequency pseudomode with respect to the linear case.

\subsection{Explicit pseudomode model for the nonlinear system-bath interaction~(\ref{eq:HintsQ})}\label{sec:PMmodel}

We now determine explicitly all unknown pseudomode parameters $\Omega_\alpha$, $\Gamma_\alpha$, $\lambda_\alpha$ in Eqs.~(\ref{eq:PMgeneralmodel}) and (\ref{eq:XPM}), so that both the two-time correlation~(\ref{eq:CXX}) and the second moment~(\ref{eq:X2}) are reproduced up to a controlled error. 

As in the analysis of Ref.~\cite{liang2024purifiedinputoutputpseudomodemodel}, we assume a spectral decomposition of the positive contribution $C(t\ge0)$ in the form of
\begin{equation}\label{eq:Cdecomppos}
C(t\ge0) \cong \sum_{l=1}^N w_l e^{-(i\nu_l+\gamma_l)t}, 
\end{equation}
with $w_l\in \mathbb{C}$, $\nu_l\in\mathbb{R}$, $\gamma_l\in\mathbb{R}_+$, and the condition $\sum_{l=1}^N w_l \cong \sigma^2$ is satisfied for consistency. Such an exponential decomposition can be realized using several existing algorithms, e.g., the adaptive Antoulas-Anderson (AAA) algorithm and the Prony decomposition, see Ref.~\cite{10.1063/5.0209348} and the references therein for a comprehensive review. Because of the time-reversal property $C(t)=C^*(-t)$, the decomposition in Eq.~(\ref{eq:Cdecomppos}) also gives the decomposition of the negative contribution $C(t\le0)$, written as
\begin{equation}\label{eq:Cdecompneg}
C(t\le0) \cong \sum_{l=1}^N w_l^* e^{-(i\nu_l-\gamma_l)t},  
\end{equation}
with $\sum_lw_l^* \cong \sigma^2$ valid by construction. 

To match the pseudomode correlation ansatz~(\ref{eq:CPM}), we combine the two decompositions in Eqs.~(\ref{eq:Cdecomppos}) and (\ref{eq:Cdecompneg}) with the representation $\Theta(\pm t) \approx e^{\pm at - a\lvert t\rvert}$ (where $a\in[0,+\infty)$ is a sufficiently large free parameter) of the Heaviside step function 
to obtain the following approximation
\begin{equation}\label{eq:CapproxCt}
C(t)\cong C_\text{pm}^a(t) = C_{\text{pm},>}^a(t) + C_{\text{pm},<}^a(t) + C_{\text{zf}}(t),
\end{equation}
in the limit $a\to+\infty$, 
with $C_{\text{zf}}(t) = - \sigma^2 e^{-a\lvert t\rvert}$ and 
\begin{equation}\label{eq:Cpnapprox}
C_{\text{pm},\gtrless}^a (t) = \sum_{l=1}^N (\text{Re}\,w_l\pm i\text{Im}\,w_l) e^{-i(\nu_l\pm ia)t - (\gamma_l+a)\lvert t\rvert}. 
\end{equation} 
The use of the symbol $\cong$ in Eq.~(\ref{eq:CapproxCt}) highlights that the approximating error is dominated by the one made using the approximation ansatz in Eq.~(\ref{eq:Cdecomppos}) and it requires further justification, which we now provide. Comparing Eq.~(\ref{eq:Cpnapprox}) with Eq.~(\ref{eq:Cdecomppos}) and Eq.~(\ref{eq:Cdecompneg}), we find that 
\begin{equation}
\begin{array}{lll}
C_{\text{pm},\gtrless}^a(t\gtrless0)&\cong& C(t\gtrless0), \\
C_{\text{pm},\gtrless}^a(t\lessgtr0)&\sim&\mathcal{O}(e^{-a\lvert t\rvert})\;,
\end{array}
\end{equation}
defining the intended forward- and backward-time nature of, respectively, $C^a_{\text{pm},>}(t)$ and $C^a_{\text{pm},<}(t)$, in the limit $a\to+\infty$.
In other words, for $t\neq0$, the overall approximation error $\lvert C(t) - C_\text{pm}^a(t)\vert$ arising from the introducing of the free parameter $a$ is of order $\mathcal{O}(e^{-a\lvert t\rvert})$, which vanishes in the limit $a\to+\infty$, justifying $C(t)\cong \lim_{a\to+\infty}C_\text{pm}^a(t)$ for $t\neq 0$.
Further analysis for the case $t=0$ is necessary due to the explicit dependence of the dynamics on $C(0)$ in the nonlinear case analyzed here, as shown at the end of  section Sec.~\ref{sec:pmtheory}. Specifically, we introduced the additional term $C_{\text{zf}}(t)$ in Eq.~(\ref{eq:CapproxCt}) to correctly reproduce the second moment of $X(0)$, as can be explicitly checked as
\begin{equation}
C_\text{pm}^a(0) = \sum_l(w_l+w_l^*) - \sigma^2 \cong 2\sigma^2-\sigma^2=\sigma^2=C(0)\;.
\end{equation}
This allows to conclude that, also at $t=0$, the error $\lvert C(t) - C_\text{pm}^a(t)\vert$ made in using  Eq.~(\ref{eq:CapproxCt}) is dominated by the spectral decomposition in Eq.~(\ref{eq:Cdecomppos}), which in principle can be made arbitrarily small. This leads to conclude that $C(t)\cong \lim_{a\to+\infty}C_\text{pm}^a(t)$ is valid for all $t\in\mathbb{R}$, ultimately justifying Eq.~(\ref{eq:CapproxCt}). 

We can now safely proceed by comparing the expressions of $C_{\text{pm},\gtrless}^a(t)$ and $C_\text{zf}(t)$ with the pseudomode correlation ansatz~(\ref{eq:CPM}), to conclude that each exponential on the right-hand side of Eq.~(\ref{eq:Cpnapprox}) can be related to a zero-temperature pseudomode labeled by the multi-index $\alpha=(l\pm)$ and specified by the 
parameters
\begin{equation}\label{eq:PMparams}
\Omega_{l\pm}=\nu_l\pm ia,~\Gamma_{l\pm}=\gamma_l+a,~\lambda_{l\pm}=\sqrt{\text{Re}w_l\pm i\text{Im}\,w_l}.
\end{equation}
Similarly, $C_\text{zf}(t)$ corresponds to one zero-frequency pseudomode characterized by the parameters
\begin{equation}\label{eq:zfparams}
\Omega_{0}= 0 ,~\Gamma_{0}=a,~\lambda_{0}=i\sigma. 
\end{equation}
The collection of these parameters defines $2N+1$ zero-temperature pseudomodes whose correlation $C_\text{pm}^a(t)$ approximates $C(t)$ in the limit $a\to+\infty$. This completes the construction of an explicit pseudomode model for general nonlinear system-bath interactions~(\ref{eq:HintsQ}). In the next section, we show that the modes constituting this model can be further purified to allow a more optimized numerical implementation of the corresponding algorithm.

\subsection{Purified pseudomode model for the nonlinear system-bath interaction (\ref{eq:HintsQ})}\label{sec:purification}

In our previous work~\cite{liang2024purifiedinputoutputpseudomodemodel} where linear system-bath interactions were considered, we have shown that under the analytical continuation in Eq.~(\ref{eq:PMparams}) for the pseudomode energy and decay parameters, the state of the pseudomodes becomes effectively pure in the limit $a\to+\infty$, leading to an optimization of the pseudomode Hilbert space which can be exploited in numerical simulations. The modes associated to these analytically-continued parameters are denoted as ``purified pseudomodes''. Here we extend the results in Ref.~\cite{liang2024purifiedinputoutputpseudomodemodel} to incorporate nonlinear interactions.

In Appendix.~\ref{sec:purificationXn}, we show that the purification technique developed in Ref.~\cite{liang2024purifiedinputoutputpseudomodemodel} can be generalized to the case $Q = X^n$. Specifically, we show that, 
in the limit $a\to+\infty$, the equation of motion in Eq.~(\ref{eq:PMgeneralmodel}) becomes the following Lindblad-like master equation 
\begin{equation}\label{eq:ppmeom}
\displaystyle i\frac{d\rho_\text{ppm}(t)}{dt} = \mathcal{L}_{S}\rho_\text{ppm} + \mathcal{L}_{0}\rho_\text{ppm} + \mathcal{L}_{\text{int},n}\rho_\text{ppm}, 
\end{equation}
where $\mathcal{L}_S[\boldsymbol\cdot] = [H_S, \boldsymbol\cdot]$ and the generators $\mathcal{L}_0$ and $\mathcal{L}_{\text{int},n}$ are defined as
\begin{equation}\label{eq:ppmLn}
\begin{array}{lll}
&\displaystyle \mathcal{L}_0[\boldsymbol\cdot] = \sum_{l=1}^N\Big[ (\nu_l-i\gamma_l)d_{l+}^\dagger d_{l+}[\boldsymbol\cdot] - (\nu_l+i\gamma_l)[\boldsymbol\cdot]d_{l-}^\dagger d_{l-} \Big], \\
&\displaystyle \mathcal{L}_{\text{int},n}[\boldsymbol\cdot] = X_+^ns[\boldsymbol\cdot] - [\boldsymbol\cdot]sX_-^n + s[\boldsymbol\cdot]X_{-,d^\dagger}^n - X_{+,d}^n[\boldsymbol\cdot]s \\
&\displaystyle~~~~~~~~~~~ + \sum_{r=1}^{n-1}\binom{n}{r} \bigg[X_+^{n-r}s[\boldsymbol\cdot]X_{-,d^\dagger}^r - X_{+,d}^r[\boldsymbol\cdot]sX_-^{n-r}\bigg].  
\end{array}
\end{equation} 
Here, $\rho_\text{ppm}(t)$ denotes the density matrix in the limit $a\to+\infty$, and we have defined the operators $X_\pm = X_{\pm,d}+X_{\pm,d^\dagger}$ with $X_{\pm,d} = \sum_{l=1}^N\lambda_{l\pm}d_{l\pm}$ and $X_{\pm,d^\dagger} = \sum_{l=1}^N\lambda_{l\pm}d_{l\pm}^\dagger$ their components to simplify the notation. The initial state is still $\rho_\text{ppm}(0)=\rho_S\otimes\rho_\text{vac}$. The generator~(\ref{eq:ppmLn}) explicitly exhibits the main feature of purified pseudomodes, namely that the annihilation and creation operators of any purified pseudomode always appear on the same side of the density matrix. Specifically, $d_{l+}$ and $d_{l+}^\dagger$ ($d_{l-}$ and $d_{l-}^\dagger$) only act on the left-hand (right-hand) side in Eq.~(\ref{eq:ppmLn}). Since all purified pseudomodes evolve from their vacuum, this critical feature directly implies that the states of all purified pseudomodes can effectively be described in terms of pure states during the evolution.  

As an important remark, we note that by taking the limit $a\to+\infty$ in Eq.~(\ref{eq:Cpnapprox}), the free correlations of the purified pseudomodes $d_{l\pm}$ can be formally identified as
\begin{equation}
C_\text{ppm}^{l\pm}(t) = (\text{Re}\,w_l\pm i\text{Im}\,w_l) e^{-i\nu_lt - \gamma_l t}\Theta(\pm t). 
\end{equation}
In addition, the error related to the free parameters $a$ in the approximation~(\ref{eq:CapproxCt}) vanishes in this limit so that the only error source of the purified pseudomode model in Eq.~(\ref{eq:ppmeom}) comes from the spectral decompositions in Eqs.~(\ref{eq:Cdecomppos}) and (\ref{eq:Cdecompneg}). As a side remark, we also note that the equation of motion in Eq.~(\ref{eq:ppmeom}) can be recast in the form of the HEOM, as shown explicitly in Appendix.~\ref{app:heom}.

We have now all the ingredients in order to obtain the purified model for the full interaction in Eq.~(\ref{eq:HintsQ}). In fact, as it can be explicitly seen in Appendix.~\ref{sec:purificationXn}, the procedure to derive the model in Eq.~(\ref{eq:ppmeom}) is additive, in the sense that for a system-bath interaction $H_{\text{int}} = sX^n + sX^m$ the generator $\mathcal{L}_\text{ppm}$ of the corresponding purified pseudomode model is simply $\mathcal{L}_\text{ppm} = \mathcal{L}_\text{S} + \mathcal{L}_0 + \mathcal{L}_{\text{int},n} + \mathcal{L}_{\text{int},m}$. This additivity arises from the fact that the coupling operator $Q_\text{pm}$ in Eq.~(\ref{eq:QPM}) is expressed as the sum of all nonlinear contributions $X_\text{pm}^n$ with weights $\alpha_n$, and the fact that the limit operation is additive. As a result, for the general nonlinear interaction~(\ref{eq:HintsQ}), the corresponding generator can be directly written as
\begin{equation}
\mathcal{L}_\text{ppm} = \mathcal{L}_S+\mathcal{L}_0 + \sum_{n=1}^\infty\alpha_n\mathcal{L}_{\text{int},n}\;,
\end{equation}
thereby concluding our construction of the purified pseudomode model.

\begin{figure}
\includegraphics[clip,width=8.5cm]
{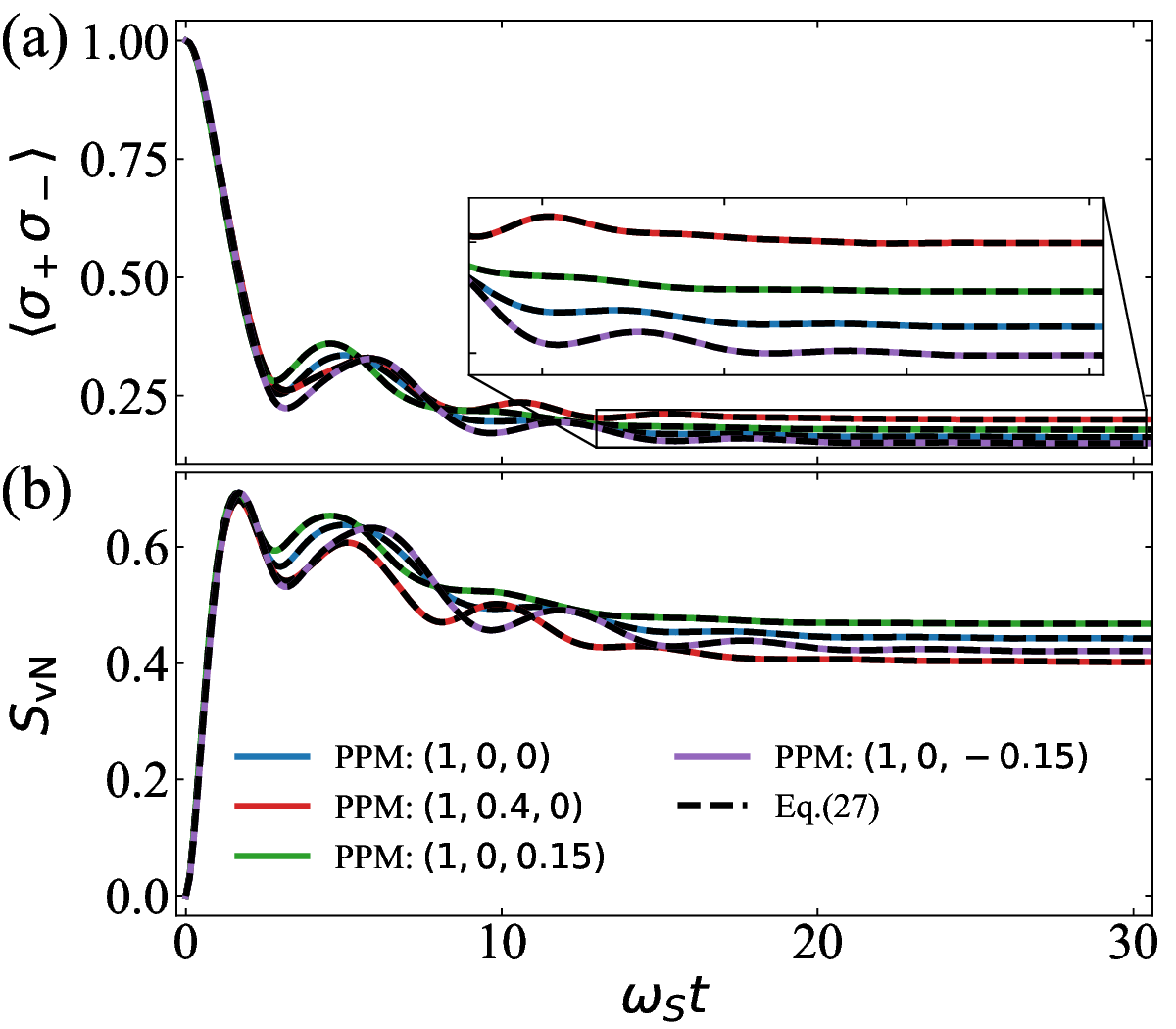}
\caption{Spontaneous decay of a two-level atom placed inside a lossy cavity. Panel (a) and (b) show the evolution of the excited state population $\langle \sigma_+\sigma_-\rangle$ and the von Neumann entropy $S_\text{vN}$, respectively. Colored solid curves are obtained using the corresponding purified pseudomode model (abbreviated as ``PPM") for four different combinations of the coefficients $(\alpha_1,\alpha_2,\alpha_3)$ in the coupling operator $Q_\text{ac} = \alpha_1X_\text{c} + \alpha_2X_\text{c}^2 + \alpha_3X_\text{c}^3$ with $X_\text{c}=\lambda(b+b^\dagger)$. Black dashed lines are the solution of the master equation in Eq.~(\ref{eq:lossymodeexample}). Simulation parameters are $\omega_S=0.5$, $\nu=0.5$, $\gamma=0.1$, and $\lambda=0.3$. 
}\label{fig:singlemode}
\end{figure}

\begin{figure}
\includegraphics[clip,width=8.5cm]
{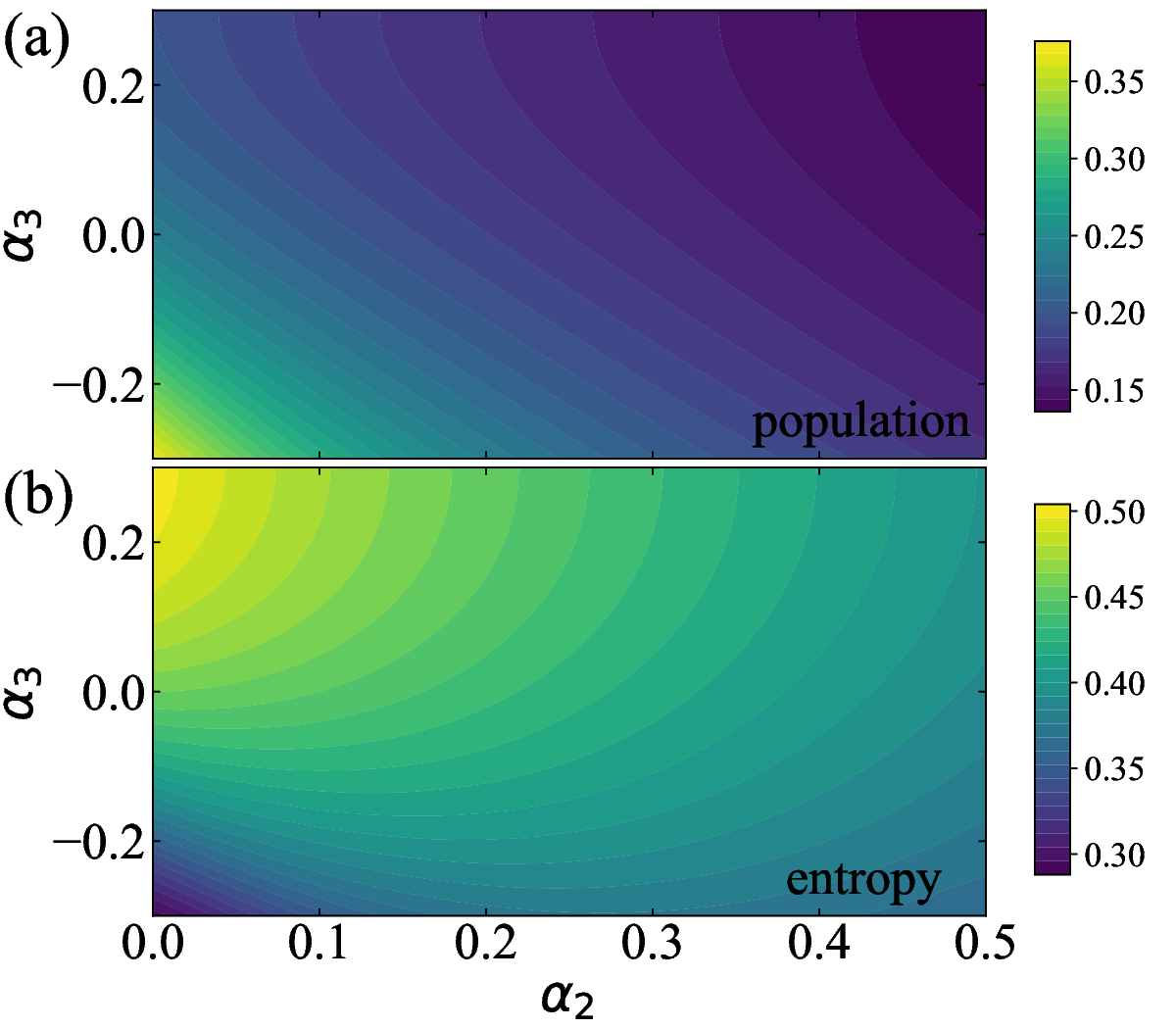}
\caption{Excited state population $\langle\sigma_+\sigma_-\rangle$ in panel (a) and von Neumann entropy $S_\text{vN}$ in panel (b) for the steady-state of the two-level atom placed in a lossy cavity. Here $\alpha_1=1$ is fixed in the coupling operator $Q_\text{ac} = \alpha_1X_\text{c} + \alpha_2X_\text{c}^2 + \alpha_3X_\text{c}^3$. Other simulation parameters are the same as those in Fig.~\ref{fig:singlemode}. 
}\label{fig:density}
\end{figure}

\begin{figure}
\includegraphics[clip,width=8.5cm]
{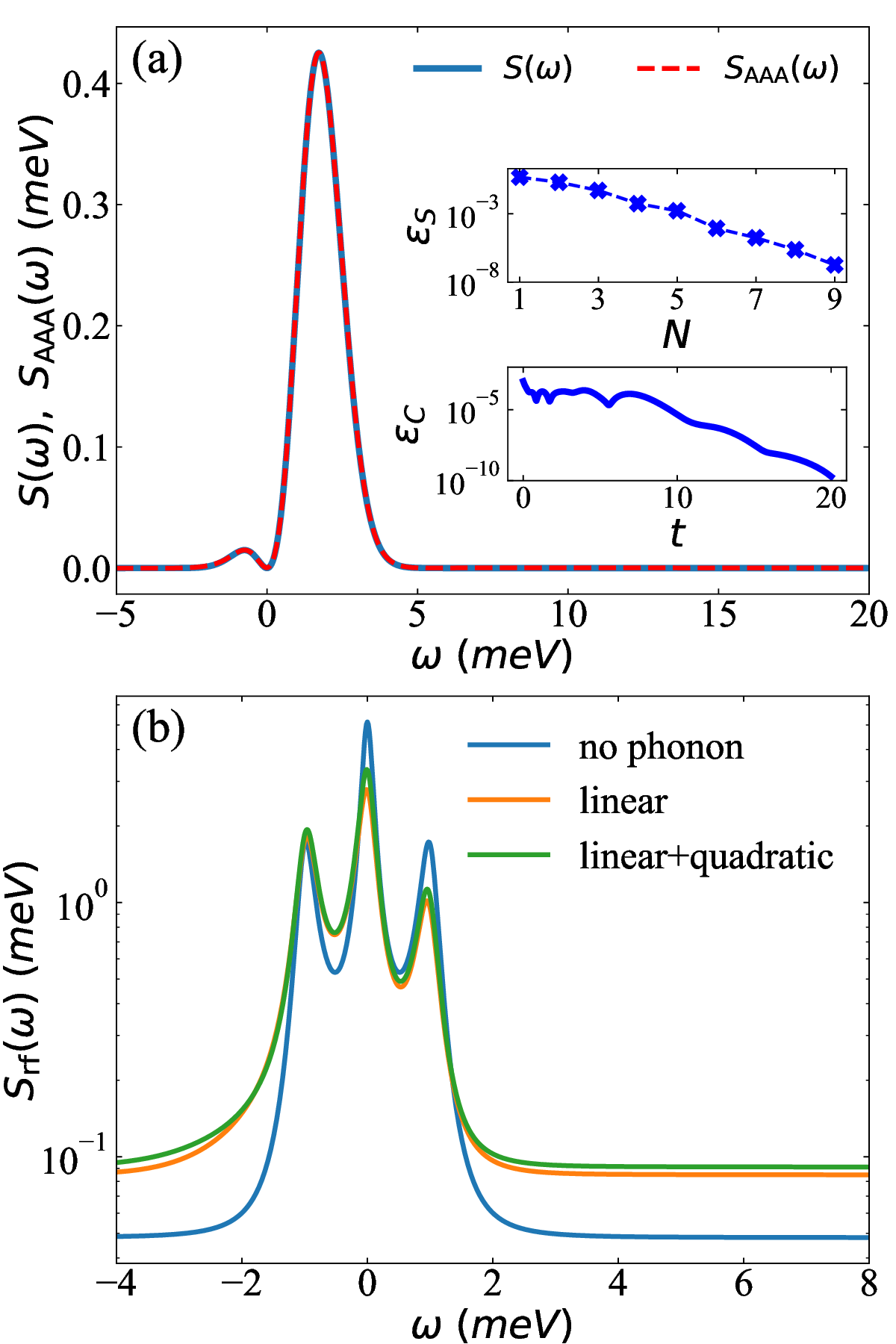}
\caption{(a) Power spectrum $S(\omega)$ (blue solid) of the phonon environment and its approximation (red dashed) generated with the AAA algorithm. The positive (negative) frequency domain of $S(\omega)$ corresponds to emission (absorption) of phonons, respectively. The upper inset shows the fitting error $\epsilon_S = \int d\omega\lvert S(\omega) - S_\text{AAA}(\omega)\rvert$ for different $N$, while the lower inset shows the fitting error $\epsilon_C = \lvert C(t) - C_\text{AAA}(t)\rvert$ as a function of time $t$. (b) Resonance fluorescence spectra in the absence (blue line) and presence (orange line for the pure linear coupling and green line for the linear+quadratic coupling) of the phonon environment. Other simulation parameters are $\kappa=\SI{0.1}{\milli\electronvolt}$ and $\Omega=\SI{1}{\milli\electronvolt}$.
}\label{fig:qd}
\end{figure}

\section{Numerical Demonstrations}\label{sec:numerics}

We proceed to illustrate the validity and potential application of our method with two examples. The numerical results presented in this section are obtained using QuTiP master equation solvers in the excitation-number restricted basis~\cite{JOHANSSON20121760,JOHANSSON20131234, Lambert2024}.

\subsection{Spontaneous decay of a two-level atom in a lossy cavity}

For benchmarking, we first consider the spontaneous decay of a two-level atom strongly coupled to a lossy cavity with the resonance frequency $\nu$ and decay rate $\gamma$, so  that the state $\rho_\text{ac}(t)$ describing both the atom and the cavity obeys the following Lindblad master equation
\begin{equation}\label{eq:lossymodeexample}
\begin{array}{lll}
\displaystyle \frac{d\rho_\text{ac}(t)}{dt} &=&\displaystyle -i[H_S+\nu b^\dagger b + sQ_\text{ac}, \rho_\text{ac}] \\
&&\displaystyle + \gamma (2b\rho_\text{ac} b^\dagger - b^\dagger b \rho_\text{ac} - \rho_\text{ac} b^\dagger b), 
\end{array}
\end{equation}
where $s=\sigma_x$, $H_S=\omega_S\sigma_z/2$ is the bare atomic Hamiltonian with level splitting $\omega_S$ and where $b$ and $b^\dagger$ denote the annihilation and creation operators of the cavity, respectively. We consider a generalized light-matter interaction consisting of both the linear term and the second- and third-order corrections. In other words, the cavity coupling operator $Q_\text{ac}$ is written as $Q_\text{ac} = \alpha_1X_\text{c} + \alpha_2X_\text{c}^2 + \alpha_3X_\text{c}^3$ with $X_\text{c}=\lambda(b+b^\dagger)$ and $\alpha_{1},\alpha_2,\alpha_3\in\mathbb{R}$. The lossy cavity can be viewed as the environment of the two-level atom and its free correlation can be solved analytically as $C(t) = \lambda^2\exp(-i\nu t - \gamma\lvert t\rvert)$. Since the bath correlation function in this case is just a single exponential, no extra spectral decomposition is needed, which is convenient to illustrate the exactness of the purification procedure described in Sec.~\ref{sec:purification}. 

The model considered here, i.e., the nonlinear open Rabi model, can be engineered in superconducting quantum circuits~\cite{PhysRevA.97.013851,PhysRevA.98.053859}. Nonlinear light-matter interactions, such as the two-photon coupling considered in Refs.~\cite{PhysRevA.97.013851,PhysRevA.98.053859,PRXQuantum.4.030326}, give rise to a number of interesting phenomena, for example, spectral collapse~\cite{PhysRevA.85.043805,PhysRevA.99.013815}, high-order optical nonlinear processes~\cite{PhysRevA.102.053710,PhysRevA.104.033701}, and novel collective photon emission~\cite{e23050612,PhysRevA.105.L011702}. They could also find applications in quantum technologies, such as quantum sensing~\cite{e24081015}, qubit-noise spectroscopy~\cite{PhysRevA.107.052603} and nonclassical state preparation~\cite{PhysRevLett.122.123604}.  

In Fig.~\ref{fig:singlemode}, we plot the evolution of the excited state population $\langle \sigma_+\sigma_-\rangle$ and atom-cavity entanglement measured by the von-Neumann entropy $S_\text{vN}=-\text{Tr}[\rho_S\ln\rho_S]$ for four different groups of the coefficients $(\alpha_1,\alpha_2,\alpha_3)$ for a resonant cavity $\nu=\omega_S$. Note that, the von-Neumann entropy is a valid entanglement measure only for zero-temperature environments~\cite{PhysRevA.53.2046}, which is the case considered here. We compare the results obtained from the purified pseudomode model (solid lines) with those from the master equation in Eq.~(\ref{eq:lossymodeexample}) (dashed black lines) and find full agreement in all cases. The presence of the small quadratic correction with $\alpha_2=0.4$ slightly increases the hybridization in the steady-state, while the the entanglement between the atom and the cavity decreases, as indicated by the red lines in Fig.~\ref{fig:singlemode}. On the other hand, by tuning the sign of the cubic correction, stronger ($\alpha_3=0.15$, green lines) or weaker ($\alpha_3=-0.15$, purple lines) light-matter hybridization and entanglement can be engineered, possibly offering an extra control knob for quantum information processing and dissipative state engineering~\cite{PhysRevResearch.6.043229}. In Fig.~\ref{fig:density} we further plot $\langle \sigma_+\sigma_-\rangle$ and $S_\text{vN}$ as functions of $\alpha_2$ and $\alpha_3$ with fixed $\alpha_1=1$ for the steady-state, which clearly reveals the opposite dependence of the excited-state population and entropy on the sign of $\alpha_3$, in stark contrast to their dependence on $\alpha_2$. Note that, we have checked that the sign of the quadratic correction has no influence on the hybridization and entanglement (not shown).

Despite the general validity of the purified pseudomode model constructed in Sec.~\ref{sec:PPMconstruction}, its numerical simulations may encounter instability issues for large nonlinearity, which arises from the necessary truncation of the infinite-dimensional Hilbert spaces of purified pseudomodes. Specifically, numerical simulations of the master equation in Eq.~(\ref{eq:ppmeom}) require truncating the infinite-dimensional pseudomode Hilbert space to a finite dimension $M$. As a consequence, the corresponding truncated generator $\mathcal{L}_\text{ppm}^{(M)}$ may give rise to eigenvalues $\xi_j^{(M)}$ with large positive imaginary part, $\text{Im}\,\xi_j^{(M)}>0$, even if the fitting error $\lvert C(t) - C_\text{pm}^\infty(t)\rvert$ is sufficiently small. Even worse, in certain circumstances $\text{Im}\,\xi_j^{(M)}$ decrease very slowly as $M$ increases. This leads to divergence in the dynamics, preventing the numerical evaluation of the steady-state using 
the truncated generator $\mathcal{L}_\text{ppm}^{(M)}$. In fact, similar issues may also occur for other pseudomode models involving unphysical model parameters and the HEOM, and several solutions to circumvent numerical instability have been proposed~\cite{10.1063/1.5092616,Krug2023} in the latter context. In our simulations, we have carefully chosen model parameters to avoid such instability. A more careful study of this practical issue will be left for future works.

\subsection{Fluorescence spectrum of quantum dots in the presence of phonon environments}
\label{sec:dot}
We next consider a more complex physical example, namely, the fluorescence spectrum of a two-level quantum dot in the presence of a phonon environment, which affects the excitation energy of the quantum dot, leading to the bath coupling operator $s=\sigma_+\sigma_-$. Unlike the previous example, the phonon environment consists of a continuum of bath modes, rendering exact results inaccessible. To measure the fluorescence spectrum, the quantum dot is driven by a resonant laser so that in the frame rotating with the laser frequency the bare Hamiltonian of the quantum dot is $H_S = \Omega\sigma_x/2$ where $\Omega$ is the Rabi freqeunce, chosen as $\SI{1}{\milli\electronvolt}$ here to be consistent with Ref.~\cite{PhysRevB.85.115309}. The quantum dot also exhibits a radiative decay with rate $\kappa$, which can be accounted for by a Lindblad dissipator with the collapse operator $\sigma_-$, i.e., $\kappa(2\sigma_-\rho_d\sigma_+ - \sigma_+\sigma_-\rho_d - \rho_d\sigma_+\sigma_-)$ where $\rho_d$ denotes the dot state. Photons emitted via this radiative channel are then collected to measure the dot fluorescence spectrum. The last ingredient of this model is the phonon environment, here characterized by the super-Ohmic spectral density $J(\omega) = \alpha_p\omega^3\exp(-\omega^2/2\omega_b^2)$, where $\alpha_p$ measures the coupling strength and $\omega_b$ is the cutoff frequency. The dot fluorescence spectrum is obtained from the Fourier transform of the two-time correlation function, i.e.,
\begin{equation}
S_\text{rf}(\omega) = \text{Re}\,\int_0^\infty d\tau e^{i\omega\tau}\big[\langle\sigma^+(\tau)\sigma^-\rangle_{ss} - \langle\sigma^+(\infty)\sigma^-\rangle_{ss} \big], 
\end{equation}
where the subscript ``ss" means the expectation is taken over the steady-state. 

In Fig.~\ref{fig:qd}(a), we plot the power spectrum $S(\omega)$ of the phonon environment, related to the spectral density $J(\omega)$ via
\begin{equation}
S(\omega) = (1+\coth(\beta\omega/2))\left[J(\omega)\theta(\omega) - J(-\omega)\theta(-\omega)\right], 
\end{equation}
where $\theta(\omega)$ is the Heaviside step function. We have chosen the parameters $\omega_b = \SI{1}{\milli\electronvolt}$, $\alpha_p/(2\pi)^2 = \SI{0.002}{\pico\second^2}$ and $T=\SI{4}{\kelvin}$ consistent with those in Ref.~\cite{PhysRevB.85.115309}. The spectral decompositions in Eqs.~(\ref{eq:Cdecomppos}) and (\ref{eq:Cdecompneg}) are generated with the AAA algorithm~\cite{doi:10.1137/16M1106122,10.1063/5.0209348}, which aims to find the rational approximation of  a general function defined on the real axis, such as $S(\omega)$, in the barycentric representation
\begin{equation}
S_\text{AAA}(\omega) = \sum_{l=1}^M \frac{z_lS(\omega_l)}{\omega - \omega_l} \Big/ \sum_{l=1}^N \frac{z_l}{\omega - \omega_l}. 
\end{equation}
Here, $M\ge1$ is an integer giving the order of the rational appproximation, $z_l\in\mathbb{C}$ are free parameters, and $\{\omega_l \in\mathbb{R}| l = 1,\cdots,M\}$ is a set of support points, over which we have $S(\omega_l) = S_\text{AAA}(\omega_l)$ using the residue theorem. After the barycentric representation $S_\text{AAA}(\omega)$ is obtained, its poles $p_l$ in the lower half complex plane and the corresponding residues $r_l$ can be solved to yield the spectral decomposition in Eq.~(\ref{eq:Cdecomppos}) with the parameters given by $w_l = -ir_l$, $\nu_l = \text{Re}\,p_l$ and $\gamma_l = -\text{Im}\,p_l$. Note that, in practice one often needs only a small portion $N$ ($N\le M$) of these poles to achieve high approximation precision. The approximate power spectrum of the phonon bath is plotted in dashed line in Fig.~\ref{fig:qd}(a). For the parameters chosen here, $N=4$ exponentials already produce a faithful approximation. 
In the two insets, we plot the fitting error $\epsilon_S = \int d\omega \lvert S(\omega) - S_\text{AAA}(\omega) \rvert$  versus $N$ (upper inset) and the fitting error $\epsilon_C = \lvert C(t) - C_\text{AAA}(t)\rvert$ as a function of time $t$ for $N=4$ (lower inset), where $C_\text{AAA}(t)$ is obtained from the Fourier transform of $S_\text{AAA}(\omega)$. As expected, $\epsilon_S$ decreases quickly when $N$ is increased while the main fitting error of $C(t)$ concentrates around $\omega=0$. A step-by-step recipe for constructing the purified pseudomode model for this example is summarized in detail in Appendix.~\ref{sec:example2}.

In Fig.~\ref{fig:qd}(b), we show the dot fluorescence spectra in the absence of the phonon environment (blue line) and in the presence of the phonon environment with a linear (orange) or linear+quadratic (green) coupling. In the first case, the dot fluorescence spectrum displays the characteristic Mollow triplet~\cite{PhysRev.188.1969} with two symmetric sidebands located around $\omega=\pm\Omega$. The emergence of sidebands can be explained in terms of laser-dressed states~\cite{PhysRev.188.1969}. A linearly coupled phonon environment makes the dot fluorescence spectrum asymmetric and slightly broadens the three peaks, as shown by the orange line in Fig.~\ref{fig:qd}(b). Asymmetric sidebands observed here arise from phonon-assisted transitions~\cite{PhysRevLett.106.247403,PhysRevB.84.085309,Hughes_2013} between laser-dressed states and can be further attributed to the asymmetry of $S(\omega)$ between phonon emission (positive frequencies) and absorption (negative frequencies) processes, see Fig.~\ref{fig:qd}(a). In the presence of a small quadratic correction, we observe an overall lift of the dot fluorescence spectrum compared to the pure linear case, see the green line in Fig.~\ref{fig:qd}(b).

\section{Conclusions and outlook}\label{sec:summary}

In summary, we have presented a generalized purified pseudomode model capable of simulating the reduced system dynamics of open quantum systems in the presence of general system-bath interactions~(\ref{eq:HintsQ}). We have pointed out that for nonlinear interactions, both the two-time bath correlation~(\ref{eq:CXX}) involving the linear coupling operator $X$ and its second moment~(\ref{eq:X2}) emerge explicitly in the Wick factorizations  of the multi-time correlation functions. We introduce an extra zero-frequency pseudomode to account for the second moment~(\ref{eq:X2}) and are able to generalize the purification technique developed in our previous work~\cite{liang2024purifiedinputoutputpseudomodemodel} to ultimately derive a purified pseudomode model for the general nonlinear system-bath interaction~(\ref{eq:HintsQ}). We have demonstrated the validity and application of our method with two numerical examples. In particular, the capability of calculating the fluorescence spectrum of quantum dots nonlinearly coupled to phonon environments could facilitate the study of phonon-assisted emission in nanophotonics. 

Let us summarize the improvement provided by our method with respect to previous studies. First, it offers a general recipe for modeling nonlinear system-bath interactions in the form of Eq. (2). This capability comes from the natural one-to-one correspondence of multi-time correlation functions between the original bath and the corresponding effective one made of pseudomodes, which avoids the combinatorial complexity in re-summing the Wick contractions required by other methods. For comparison, Refs.~\cite{10.1063/1674-0068/30/cjcp1706123,10.1063/1.4991779} analyzed a similar problem within the HEOM formalism, 
but restricted to the case of linear interaction plus quadratic corrections, where the Wick re-summation is still tractable. Second, our method is non-perturbative and accounts for the multi-mode nature of bosonic quantum environments. In contrast, Ref.~\cite{PhysRevLett.132.106001} simplified the description of the phonon bath by replacing it with a single, most relevant, bath mode, while Refs.~\cite{PhysRevA.97.013851,e23050612,PRXQuantum.4.030326} made use of the rotating wave approximation valid only for weak interaction strength in the context of cavity or waveguide QED. Our method thus paves the way for an accurate description of the effects of nonlinear light- and phonon-matter interactions. Finally, purification of the pseudomode state leads to a reduction in the local dimension of the state space, which can be exploited in tensor-network simulations to reduce computational complexity. This has been  demonstrated in our previous work~\cite{liang2024purifiedinputoutputpseudomodemodel} for linear system-bath interactions.

As an outlook, the method developed in this work could be applied to the modeling of other types of nonlinear system-bath interactions, such as the one in Ref.~\cite{PRXQuantum.4.030326} engineered in waveguide QED and the Peierls-type light-matter interactions realized in cavity quantum materials~\cite{10.1063/5.0083825,PhysRevLett.131.023601,1lpw-22np}. Possible further extensions of this method include the modeling of nonlinear interactions in fermionic environments~\cite{PhysRevResearch.5.033011}, the computation of bath properties~\cite{liang2024purifiedinputoutputpseudomodemodel,4q4p-z14x}, and the incorporation of stochastic techniques~\cite{PRXQuantum.4.030316} and tensor-network algorithms~\cite{SCHOLLWOCK201196,PhysRevB.94.165116,PhysRevX.10.031040}.

\begin{acknowledgments}
We would like to thank Dr. W. Yang for interesting discussions about the role of nonlinearities in the pseudomode model. M.C. acknowledges support from NSFC (Grant No. 12088101) and NSAF (Grant No. U2330401).
\end{acknowledgments}

\appendix

\section{Detailed derivations for the purification of the pseudomode model}\label{sec:purificationXn}

In this section, we show that the purification technique first presented in our previous work~\cite{liang2024purifiedinputoutputpseudomodemodel} can be generalized to the case $Q=X^n$, a special case of the general nonlinear interaction~(\ref{eq:HintsQ}). 

We start by rewriting the pseudomode model given by Eqs.~(\ref{eq:PMgeneralmodel}), (\ref{eq:QPM}),  (\ref{eq:PMparams}) and (\ref{eq:zfparams}) in the Liouville space 
\begin{equation}
\begin{array}{lll}
\mathcal{L} &\displaystyle = \mathcal{H}_S\otimes\mathcal{H}_S\otimes
\Big(\bigotimes_{l=1}^N\big(\mathcal{H}_{\text{pm},l+}\otimes\mathcal{H}_{\text{pm},l+}\big)\otimes\mathcal{H}_0\Big)  \\
&\displaystyle~~~\otimes\Big(\bigotimes_{l=1}^N\big(\mathcal{H}_{\text{pm},l-}\otimes\mathcal{H}_{\text{pm},l-}\big)\otimes\mathcal{H}_0\Big), 
\end{array}
\end{equation}
where $\mathcal{H}_S$, $\mathcal{H}_{\text{pm},l\pm}$ and $\mathcal{H}_0$ denote the Hilbert space of the system, the pseudomode indexed by $\alpha=l\pm$ and the zero-frequency pseudomode, respectively.  To achieve this, 
we employ the ``bra-flipper" superoperator $\mathfrak{U}$ introduced in Ref.~\cite{Gyamfi_2020} to define the mapping
\begin{equation}
\mathfrak{U}[|\psi\rangle \langle \psi'|] = |\psi\rangle\otimes|\psi'\rangle^* \equiv |\psi\psi'\rangle\rangle, 
\end{equation}
for two state vectors $|\psi\rangle$ and $|\psi'\rangle$ in a Hilbert space. 
We denote orthonormal basis states of the system $S$ as $|s\rangle$ and $|s'\rangle$, and denote  Fock states corresponding to the pseudomode indexed by $\alpha=l\pm$ ($\alpha=0$) as $|n_{l\pm}\rangle$ and $|n_{l\pm}'\rangle$ ($|n_0\rangle$ and $|n_0'\rangle$). In this basis, the density matrix $\rho_\text{pm}$ in Eq.(\ref{eq:PMgeneralmodel}) can be expanded as 
\begin{equation}
\begin{array}{lll}
&\displaystyle \rho_\text{pm} = \sum_{s,s',n_{l\pm},n_{l\pm}'} (\rho_\text{pm})_{sn_{l+}n_{l-},s'n_{l+}'n_{l-}'} \times \\
&\displaystyle |s\rangle\langle s'|\otimes\Big(\bigotimes_{l=1}^N\big(|n_{l+}\rangle\langle n_{l+}'|\otimes|n_{l-}\rangle\langle n_{l-}'|\big)\Big)\otimes|n_0\rangle\langle n_0'|.  
\end{array}
\end{equation}   
Applying the bra-flipper superoperator $\mathfrak{U}$, we can vectorize $\rho_\text{pm}$ as 
\begin{equation}
\begin{array}{lll}
&\displaystyle |\rho_\text{pm}\rangle\rangle =  \sum_{s,s',n_{l\pm},n_{l\pm}'} (\rho_\text{pm})_{sn_{l+}n_{l-},s'n_{l+}'n_{l-}'} \times \\
&\displaystyle |ss'\rangle\rangle\otimes\Big(\bigotimes_{l=1}^N\big(|n_{l+}n_{l+}'\rangle\rangle\otimes|n_{l-}n_{l-}'\rangle\rangle\big)\Big)\otimes|n_0n_0'\rangle\rangle.  
\end{array}
\end{equation}   
On the other hand, this mapping of the density matrix also gives rise to the mapping of the left (right) action of 
an operator $O$, with support on either $\mathcal{H}_\text{S}$ or $\mathcal{H}_{\text{PM},l\pm}$ or $\mathcal{H}_0$,  to $\bar{O}\equiv O\otimes I$ ($\tilde{O} \equiv I\otimes O^\intercal$), 
with support on  either $\mathcal{H}_\text{S}\otimes\mathcal{H}_\text{S}$ or $\mathcal{H}_{\text{pm},l\pm}\otimes\mathcal{H}_{\text{pm},l\pm}$ or $\mathcal{H}_0\otimes\mathcal{H}_0$. Here, $I$ denotes the corresponding identity operator and $O^\intercal$ the transpose of $O$. 

Ultimately, in the Liouville space, the equation of motion in Eq.~(\ref{eq:PMgeneralmodel}) can be rewritten as
\begin{equation}
\begin{array}{lll}
\displaystyle i\frac{d|\rho_\text{pm}^a\rangle\rangle}{dt} = \mathfrak{L}|\rho_\text{pm}^a\rangle\rangle = (\mathfrak{L}_S+\mathfrak{L}_{\text{pm},0}^a+\mathfrak{L}_{\text{PM},\text{int}}^a)|\rho_\text{pm}^a\rangle\rangle, 
\end{array}
\end{equation}
with $\mathfrak{L}_\text{S} = \bar{H}_\text{S} - \tilde{H}_\text{S}^\intercal$ and
\begin{equation}
\begin{array}{lll}
\displaystyle \mathfrak{L}_{\text{pm},0}^a &\displaystyle = \sum_{l=1}^N\Big[ (\nu_{l}-i\gamma_{l})\bar{d}_{l+}^\dagger\bar{d}_{l+} - (\nu_{l}+i(\gamma_{l}+2a))\tilde{d}_{l+}^\dagger\tilde{d}_{l+} \\
&\displaystyle~~~ + 2i(\gamma_{l}+a)\bar{d}_{l+}\tilde{d}_{l+} + (\nu_l-i(\gamma_l+2a))\bar{d}_{l-}^\dagger\bar{d}_{l-}   \\
&\displaystyle~~~  - (\nu_l+i\gamma_l)\tilde{d}_{l-}^\dagger\tilde{d}_{l-} + 2i(\gamma_l+a)\bar{d}_{l-}\tilde{d}_{l-}   \\
&\displaystyle~~~ + 2ia\bar{d}_0\tilde{d}_0 - ia\bar{d}_0^\dagger\bar{d}_0 - ia\tilde{d}_0^\dagger\tilde{d}_0 \Big],  
\end{array}
\end{equation}   
and 
\begin{equation}
\displaystyle \mathfrak{L}_{\text{PM},\text{int}}^a = \bar{s} \Big[ \bar{X}_+ + \bar{X}_- + \bar{X}_0 \Big]^n - \tilde{s}^\intercal \Big[\tilde{X}_+ + \tilde{X}_- + \tilde{X}_0 \Big]^n, 
\end{equation}   
where for later convenience we have introduced the operators $\bar{X}_\xi = \bar{X}_{\xi,d} + \bar{X}_{\xi,d^\dagger}$, $\tilde{X}_\xi = \tilde{X}_{\xi,d} + \tilde{X}_{\xi,d^\dagger}$ with $\xi\in\{+,-,0\}$, whose components are $\bar{X}_{\pm,d} = \sum_{l=1}^N\lambda_{l\pm}\bar{d}_{l\pm}$, $\bar{X}_{\pm,d^\dagger} = \sum_{l=1}^N\lambda_{l\pm}\bar{d}_{l\pm}^\dagger$, $\tilde{X}_{\pm,d} = \sum_{l=1}^N\lambda_{l\pm}\tilde{d}_{l\pm}$, $\tilde{X}_{\pm,d^\dagger} = \sum_{l=1}^N\lambda_{l\pm}\tilde{d}_{l\pm}^\dagger$, $\bar{X}_{0,d} = \lambda_0 \bar{d}_0$, $\bar{X}_{0,d^\dagger} = \lambda_0 \bar{d}_0^\dagger$, $\tilde{X}_{0,d} = \lambda_0 \tilde{d}_0$ and $\tilde{X}_{0,d^\dagger} = \lambda_0 \tilde{d}_0^\dagger$, respectively. In addition, the initial state $\rho_\text{pm}(0)$ is now mapped to the vector
\begin{equation}
|\rho_\text{pm}(0)\rangle\rangle = |\rho_\text{S}(0)\rangle\rangle\otimes\bigotimes_{l=1}^N|\bar0_{l+}\tilde0_{l+}\bar0_{l-}\tilde0_{l-}\rangle\rangle\otimes|\bar{0}_0\tilde{0}_0\rangle\rangle,    
\end{equation}
where the vacuum states of the modes $\bar{d}_{l\pm/0}$ and $\tilde{d}_{l\pm/0}$ are denoted as $|\bar{0}_{l\pm/0}\rangle$ and $|\tilde{0}_{l\pm/0}\rangle$, respectively.  

To gain more intuition, we reiterate the argument in the previous work~\cite{liang2024purifiedinputoutputpseudomodemodel}. We observe that the decay rates of the modes $\tilde{d}_{l+}$, $\bar{d}_{l-}$, $\tilde{d}_0$ and $\bar{d}_0$ depend on the free parameter $a$, meaning that in the limit $a\to+\infty$ these modes decay so fast that they should be frozen in their initial vacuum states. This implies that in this limit these modes can be adiabatically eliminated so that they are  dynamically decoupled from the other degrees of freedom. 

To confirm this intuitive picture, we expand the interaction Liouville operator $\mathfrak{L}_{\text{PM},\text{int}}^a$ as
\begin{equation}\label{eq:Lint1}
\begin{array}{lll}
\mathfrak{L}_{\text{PM},\text{int}}^a &\displaystyle = \bar{s} \sum_{r=0}^n \binom{n}{r} \bar{X}_{+}^{n-r}\bar{X}_{-0}^r - \tilde{s}^\intercal \sum_{r=0}^n \binom{n}{r} \tilde{X}_{+0}^r \tilde{X}_-^{n-r} \\
&\displaystyle = \bar{s} \sum_{r=0}^n\sum_{p=0}^r \binom{n}{r}\binom{r}{p} \bar{X}_{+}^{n-r}\bar{X}_{-0,d^\dagger}^p\bar{X}_{-0,d}^{r-p} \\
&\displaystyle~~~ - \tilde{s}^\intercal \sum_{r=0}^n\sum_{p=0}^r \binom{n}{r}\binom{r}{p}\tilde{X}_-^{n-r}\tilde{X}_{+0,d^\dagger}^p\tilde{X}_{+0,d}^{r-p}, 
\end{array}
\end{equation}
where in the first line we have introduced the shorthand notations $\bar{X}_{-0} = \bar{X}_- + \bar{X}_0$ and $\tilde{X}_{-0} = \tilde{X}_- + \tilde{X}_0$, and in the second line we have used the fact that the component $\bar{X}_{-0,d}$ ($\tilde{X}_{+0,d}$) commutes with $\bar{X}_{-0,d^\dagger}$ ($\tilde{X}_{+0,d^\dagger}$), i.e., 
\begin{equation}
\begin{array}{lll}
[\bar{X}_{-0,d}, \bar{X}_{-0,d^\dagger}] &= \sum_{l=1}^N\lambda_{-}^2 + \lambda_0^2 = 0,  \\
{[\tilde{X}_{+0,d}, \tilde{X}_{+0,d^\dagger}]} &= \sum_{l=1}^N\lambda_{+}^2 + \lambda_0^2 = 0, 
\end{array}
\end{equation}
since the equalities $\sum_{l=1}^N\lambda_{l-}^2 = \sum_{l=1}^Nw_l^* = \sigma^2$, $\sum_{l=1}^N\lambda_{l+}^2 = \sum_{l=1}^Nw_l = \sigma^2$ and $\lambda_0^2 = -\sigma^2$ are valid by construction, see Eqs.~(\ref{eq:Cdecomppos}), (\ref{eq:Cdecompneg}) and (\ref{eq:zfparams}). Based on Eq.~(\ref{eq:Lint1}), we further classify the terms in the interaction Liouville operator $\mathfrak{L}_{\text{PM},\text{int}}^a$ as 
\begin{equation}\label{eq:Lint2}
\mathfrak{L}_{\text{PM},\text{int}}^a = \mathfrak{V} + \big(\bar{s} \bar{X}_{+}^{n} - \tilde{s}^\intercal \tilde{X}_-^{n}\big) + \mathfrak{R}, 
\end{equation}
with 
\begin{equation}\label{eq:Vterm}
\begin{array}{lll}
\displaystyle\mathfrak{V} &=&\displaystyle \bar{s} \sum_{r=1}^n \binom{n}{r} \bar{X}_{+}^{n-r} \bigg[ \sum_{p=0}^r\binom{r}{p} \bar{X}_{-,d^\dagger}^p\bar{X}_{0,d^\dagger}^{r-p} \bigg] \\
&&\displaystyle - \tilde{s}^\intercal \sum_{r=1}^n \binom{n}{r}\tilde{X}_-^{n-r} \bigg[ \sum_{p=0}^r\binom{r}{p}\tilde{X}_{+,d^\dagger}^p\tilde{X}_{0,d^\dagger}^{r-p} \bigg],  
\end{array}
\end{equation}
and 
\begin{equation}\label{eq:Rterm}
\begin{array}{lll}
\mathfrak{R} &\displaystyle= \bar{s} \sum_{r=1}^n\sum_{p=0}^{r-1} \binom{n}{r}\binom{r}{p} \bar{X}_+^{n-r} \bar{X}_{-0,d^\dagger}^p \bar{X}_{-0,d}^{r-p} \\
&\displaystyle~~~ - \tilde{s}^\intercal \sum_{r=1}^n\sum_{p=0}^{r-1} \binom{n}{r}\binom{r}{p} \tilde{X}_-^{n-r} \tilde{X}_{+0,d^\dagger}^p \tilde{X}_{+0,d}^{r-p}. 
\end{array}
\end{equation}
The reason underlying this classification is that for the modes with enhanced dissipation, namely, $\tilde{d}_{l+}$, $\bar{d}_{l-}$, $\tilde{d}_0$ and $\bar{d}_0$, the term $\mathfrak{V}$ in Eq.~(\ref{eq:Vterm}) only involves their creation operators $\bar{X}_{-,d^\dagger}$, $\tilde{X}_{+,d^\dagger}$, $\bar{X}_{0,d^\dagger}$ and $\tilde{X}_{0,d^\dagger}$, meaning that $\mathfrak{V}$ can trigger transitions out of the vacuum of $\tilde{d}_{l+}$, $\bar{d}_{l-}$, $\tilde{d}_0$ and $\bar{d}_0$. Hence, the term $\mathfrak{V}$ should be eliminated to justify the aforementioned intuitive argument. In contrast, the term $\mathfrak{R}$ in Eq.~(\ref{eq:Rterm}) contains operator strings ending with either $\bar{X}_{-0,d}$ or $\tilde{X}_{+0,d}$. If $\mathfrak{V}$ can be adiabatically  eliminated, the term $\mathfrak{R}$ has no influence on the dynamics, therefore can be simply dropped from the generator $\mathfrak{L}$. 

We can now generalize the technique presented in the previous work~\cite{liang2024purifiedinputoutputpseudomodemodel} to adiabatically eliminate the term $\mathfrak{V}$ from the generator $\mathfrak{L}=\mathfrak{L}_S+\mathfrak{L}_{\text{pm},0}^a+\mathfrak{L}_{\text{PM},\text{int}}^a$. In particular, we look for a Liouvillian operator $\mathfrak{S}$, which, crucially, we require to be $\mathcal{O}(a^{-1})$, to define the Schrieffer-Wolff (SW) transformation 
\begin{equation}\label{eq:SW}
e^{\mathfrak{S}}\mathfrak{L}e^{-\mathfrak{S}} = \mathfrak{L} + [\mathfrak{S},\mathfrak{L}] + \frac{1}{2!}[\mathfrak{S},[\mathfrak{S},\mathfrak{L}]] + \cdots, 
\end{equation}
such that the following operator equation is satisfied
\begin{equation}\label{eq:Seq}
[\mathfrak{S}, \mathfrak{L}_0] = -\mathfrak{V} + \mathfrak{V}'+\mathfrak{V}_\text{res}+\mathcal{O}(a^{-1}), 
\end{equation}
with $\mathfrak{L}_0 = \mathfrak{L}_S + \mathfrak{L}_{\text{pm},0}^a + \big(\bar{s} \bar{X}_{+}^{n} - \tilde{s}^\intercal \tilde{X}_-^{n}\big) + \mathfrak{R}$. Here, the operator $\mathfrak{V}'$ represents the emerging mediated interactions, and for consistency we require, not to contain any operators related to the modes $\tilde{d}_{l+}$, $\bar{d}_{l-}$, $\tilde{d}_0$ and $\bar{d}_0$, and to be $\mathcal{O}(a^0)$. The operator $\mathfrak{V}_\text{res}$ contains terms ending with either $\tilde{X}_{+,d}$ or $\bar{X}_{-,d}$ or $\tilde{X}_{0,d}$ or $\bar{X}_{0,d}$, which, for the same reason concerning the term $\mathfrak{R}$ in Eq.(\ref{eq:Lint2}) renders $\mathfrak{V}_\text{res}$ irrelevant to the dynamics. Importantly, the existence of the commutator in Eq.~(\ref{eq:Seq}) ensures nested commutators appearing in the SW transformation~(\ref{eq:SW}), such as  $[\mathfrak{S},\cdots,[\mathfrak{S},\mathfrak{L}_0]]$ with at least two $\mathfrak{S}$, or $[\mathfrak{S},\cdots,[\mathfrak{S},\mathfrak{V}]]$ with at least one $\mathfrak{S}$, to be at least of order $\mathcal{O}(a^{-1})$. Hence, in the limit $a\to+\infty$, they vanish in the expansion of the SW transformation~(\ref{eq:SW}), leading to the simple generator 
\begin{equation}\label{eq:SLS}
e^{\mathfrak{S}}\mathfrak{L}e^{-\mathfrak{S}} = \mathfrak{L}_0 + \mathfrak{V}' + \mathfrak{V}_\text{res}, 
\end{equation}
under which the states of the modes $\tilde{d}_{l+}$, $\bar{d}_{l-}$, $\tilde{d}_0$ and $\bar{d}_0$ to be eliminated indeed remain on their initial vacuum states. Ultimately, this allows to discard $\mathfrak{R}$ and $\mathfrak{V}_\text{res}$ from Eq.~(\ref{eq:SLS}), leading to an effective generator 
\begin{equation}\label{eq:newGen}
\begin{array}{lll}
\displaystyle \mathfrak{L}_n &=&\displaystyle \mathfrak{L}_S + \sum_{l=1}^N \Big[ (\nu_{l}-i\gamma_{l})\bar{d}_{l+}^\dagger\bar{d}_{l+} - (\nu_l+i\gamma_l)\tilde{d}_{l-}^\dagger\tilde{d}_{l-} \Big] \\
&&\displaystyle + \bar{s} \bar{X}_{+}^{n} - \tilde{s}^\intercal \tilde{X}_-^{n} +\mathfrak{V}'. 
\end{array}
\end{equation}
In addition, since $\mathfrak{S}$ is of order $\mathcal{O}(a^{-1})$, we have $\lim_{a\to+\infty} e^{\mathfrak{S}}|\rho_\text{pm}\rangle\rangle = |\rho_\text{pm}\rangle\rangle$, meaning that in the limit $a\to+\infty$ the vectorized density matrix remains invariant under the SW transformation. 

The remaining task is to solve the operator equation in Eq.~(\ref{eq:Seq}). We have found the solution
\begin{equation}\label{eq:Ssol}
\begin{array}{lll}
&&\displaystyle -2ia\mathfrak{S} \\
&=&\displaystyle \bar{s} \Bigg[ \sum_{p=0}^{n-1} \frac{1}{n-p}\bar{X}_{-,d^\dagger}^{n-p}\tilde{X}_{-,d}^p \\
&&\displaystyle~~~ + \sum_{r=1}^{n-1}\sum_{p=0}^{r-1} \frac{1}{r-p}\binom{n}{r} \bar{X}_{-,d^\dagger}^{r-p}\tilde{X}_{-,d}^p \bar{X}_+^{n-r} \Bigg] \\
&&\displaystyle - \tilde{s}^\intercal \Bigg[ \sum_{p=0}^{n-1} \frac{1}{n-p}\tilde{X}_{+,d^\dagger}^{n-p}\bar{X}_{+,d}^p \\
&&\displaystyle~~~~~~~ + \sum_{r=1}^{n-1}\sum_{p=0}^{r-1} \frac{1}{r-p}\binom{n}{r} \tilde{X}_{+,d^\dagger}^{r-p}\bar{X}_{+,d}^p \tilde{X}_-^{n-r}  \Bigg] \\
&&\displaystyle + \sum_{r=1}^n \frac2r\binom{n}{r} \bigg[ \bar{s}\bar{X}_+^{n-r}\bar{X}_{0,d^\dagger}^r - \tilde{s}^\intercal\tilde{X}_{-}^{n-r}\tilde{X}_{0,d^\dagger}^r \bigg] \\
&&\displaystyle + \bar{s} \sum_{r=1}^n \binom{n}{r}\bar{X}_+^{n-r} \sum_{p=1}^{r-1}\binom{r}{p}\frac{2p}{r+p} \\
&&\displaystyle \times \bigg[ \sum_{m=0}^{p-1}A_m \bar{X}_{-,d^\dagger}^{p-m}\bar{X}_{0,d^\dagger}^{r-p}\tilde{X}_{-,d}^m + \frac{2A_{p-1}}{r-p} \bar{X}_{0,d^\dagger}^{r-p}\tilde{X}_{-,d}^{p} \bigg] \\
&&\displaystyle - \tilde{s}^\intercal \sum_{r=2}^n \binom{n}{r}\tilde{X}_-^{n-r} \sum_{p=1}^{r-1}\binom{r}{p}\frac{2p}{r+p} \\
&&\displaystyle \times \bigg[ \sum_{m=0}^{p-1}A_m \tilde{X}_{+,d^\dagger}^{p-m}\tilde{X}_{0,d^\dagger}^{r-p}\bar{X}_{+,d}^m + \frac{2A_{p-1}}{r-p}\tilde{X}_{0,d^\dagger}^{r-p}\bar{X}_{+,d}^p  \bigg], 
\end{array}
\end{equation}
where the coefficients $A_m$, $m\in\mathbb{N}$, are defined as $A_0 = 1/p$ and for $m\ge1$ 
\begin{equation}
A_m = \frac1p\prod_{k=1}^{m} \frac{2(p-k+1)}{r+p-2k},
\end{equation}
satisfying the recurrence relation $A_m(r+p-2m)/2 = A_{m-1}(p-m+1)$. Substituting the solution in Eq.~(\ref{eq:Ssol}) into Eq.~(\ref{eq:Seq}) leads to the new mediated interaction 
\begin{equation}\label{eq:Vprime}
\begin{array}{lll}
\mathfrak{V}' &=&\displaystyle \bar{s} \bigg[ \tilde{X}_{-,d}^n + \sum_{r=1}^{n-1}\binom{n}{r} \bar{X}_+^{n-r}\tilde{X}_{-,d}^r \bigg]  \\
&&\displaystyle - \tilde{s}^\intercal \bigg[ \bar{X}_{+,d}^n+ \sum_{r=1}^{n-1}\binom{n}{r} \bar{X}_{+,d}^r\tilde{X}_-^{n-r} \bigg], \\
\end{array}
\end{equation}
and the residual term
\begin{equation}\label{eq:Vres}
\begin{array}{lll}
&&\mathfrak{V}_\text{res} \\
&=&\displaystyle \bar{s}\sum_{r=1}^{n-1}\sum_{p=0}^{r-1} \frac{n-r}{r-p} \binom{n}{r}\bar{X}_{-,d^\dagger}^{r-p}\tilde{X}_{-,d}^p \bar{X}_+^{n-r-1} \tilde{X}_{+,d} \\
&&\displaystyle - \tilde{s}^\intercal \sum_{r=1}^{n-1}\sum_{p=0}^{r-1}\frac{n-r}{r-p}\binom{n}{r} \tilde{X}_{+,d^\dagger}^{r-p}\bar{X}_{+,d}^p\tilde{X}_{-}^{n-r-1}\bar{X}_{-,d}  \\
&&\displaystyle + 2\sum_{r=1}^n \binom{n}{r} \bigg[ \bar{s}\bar{X}_+^{n-r}\bar{X}_{0,d^\dagger}^{r-1}\tilde{X}_{0,d} - \tilde{s}^\intercal\tilde{X}_{-}^{n-r}\tilde{X}_{0,d^\dagger}^{r-1}\bar{X}_{0,d}\bigg] \\
&&\displaystyle + \bar{s} \sum_{r=2}^n \binom{n}{r} \sum_{p=1}^{r-1}\binom{r}{p} \frac{2p}{r+p}\sum_{m=1}^{p-1}A_m\bar{X}_{-,d^\dagger}^{p-m}\tilde{X}_{-,d}^m \\
&&\displaystyle \times \bigg[ (r-p)\bar{X}_+^{n-r}\bar{X}_{0,d^\dagger}^{r-p-1}\tilde{X}_{0,d} \\
&&\displaystyle~~~~ + (n-r)\bar{X}_+^{n-r-1}\bar{X}_{0,d^\dagger}^{r-p}\tilde{X}_{+,d} \bigg]  \\
&&\displaystyle - \tilde{s}^\intercal \sum_{r=2}^n \binom{n}{r} \sum_{p=1}^{r-1}\binom{r}{p}\frac{2p}{r+p} \sum_{m=0}^{p-1}A_m \tilde{X}_{+,d^\dagger}^{p-m}\bar{X}_{-,d}^m \\
&&\displaystyle \times \bigg[  (r-p) \tilde{X}_-^{n-r}\tilde{X}_{0,d^\dagger}^{r-p-1}\tilde{X}_{0,d} \\
&&\displaystyle~~~~ +  (n-r)\tilde{X}_-^{n-r-1}\tilde{X}_{0,d^\dagger}^{r-p}\bar{X}_{-,d} \bigg] \\
&&\displaystyle + \bar{s} \sum_{r=2}^n \binom{n}{r} \sum_{p=1}^{r-1}\binom{r}{p}\frac{4p}{r+p}A_{p-1}\tilde{X}_{-,d}^{p} \\
&&\displaystyle \times \bigg[ \bar{X}_+^{n-r}\bar{X}_{0,d^\dagger}^{r-p-1}\tilde{X}_{0,d} + \frac{n-r}{r-p}\bar{X}_+^{n-r-1}\bar{X}_{0,d^\dagger}^{r-p}\tilde{X}_{+,d} \bigg] \\
&&\displaystyle - \tilde{s}^\intercal \sum_{r=2}^n \binom{n}{r} \sum_{p=1}^{r-1}\binom{r}{p}\frac{4p}{r+p}A_{p-1}\bar{X}_{+,d}^{p} \\
&&\displaystyle\times \bigg[ \tilde{X}_-^{n-r}\tilde{X}_{0,d^\dagger}^{r-p-1}\bar{X}_{0,d} + \frac{n-r}{r-p}\tilde{X}_-^{n-r-1}\tilde{X}_{0,d^\dagger}^{r-p}\bar{X}_{-,d} \bigg].   
\end{array}
\end{equation}
Switching back to the density matrix representation, we arrive at the superoperators in Eq.~(\ref{eq:ppmLn}).  

\section{Recasting Eq.~(\ref{eq:ppmeom}) into the hierarchical equations of motion}\label{app:heom}

To rewrite Eq.~(\ref{eq:ppmeom}) in the form of the HEOM, we denote Fock states of purified pseudomodes as $|\mathbf{m},\mathbf{n}\rangle$, where the sequences $\mathbf{m}=(m_1,m_2,\cdots)$ and $\mathbf{n}=(n_1,n_2,\cdots)$ specify the occupations $m_l$ and $n_l$ of the modes $d_{l\pm}$, respectively. As pointed out in the main text, in Eq.~(\ref{eq:ppmLn}) the annihilation and creation operators of a purified pseudomode always appear on the same side of the density matrix. This observation combined with the assumption that all purified pseudomdoes are initialized in their vacuum allow us to conclude that only the matrices $\rho_{\mathbf{m},\mathbf{n}}\equiv\langle\mathbf{m},\mathbf{0}|\rho_\infty|\mathbf{0},\mathbf{n}\rangle$ are nonvanishing, which correspond to the auxiliary density matrices (ADOs) in the HEOM. These ADOs have the same dimensions as the reduced density matrix $\rho_{S}$. According to Eq.~(\ref{eq:ppmeom}), the ADOs $\rho_{\mathbf{m},\mathbf{n}}$ obey the equations of motion 
\begin{equation}\label{eq:heom}
\begin{array}{lll}
&\displaystyle id\rho_{\mathbf{m},\mathbf{n}}/dt \\
&\displaystyle = [H_S,\rho_{\mathbf{m},\mathbf{n}}] \\
&\displaystyle + \sum_l\Big((\nu_l-i\gamma_l)m_l\rho_{\mathbf{m},\mathbf{n}} - (\nu_l+i\gamma_l)n_l\rho_{\mathbf{m},\mathbf{n}} \Big)\\
&\displaystyle + \sum_{\mathbf{m}'} \langle\mathbf{m}|X_+^n|\mathbf{m}'\rangle s\rho_{\mathbf{m}',\mathbf{n}} - \sum_{\mathbf{n}'} \langle\mathbf{n}'|X_-^n|\mathbf{n}\rangle \rho_{\mathbf{m},\mathbf{n}'}s \\
&\displaystyle + \sum_{\mathbf{n}'} \langle\mathbf{n}'|X_{-,d^\dagger}^n|\mathbf{n}\rangle s\rho_{\mathbf{m},\mathbf{n}'} - \sum_{\mathbf{m}'} \langle\mathbf{m}|X_{+,d}^n|\mathbf{m}'\rangle \rho_{\mathbf{m}',\mathbf{n}}s \\
&\displaystyle + \sum_{r=1}^{n-1}\binom{n}{r}\sum_{\mathbf{m}',\mathbf{n}'} \bigg[  \langle\mathbf{m}|X_+^{n-r}|\mathbf{m}'\rangle\langle\mathbf{n}'|X_{-,d^\dagger}^{n-r}|\mathbf{n}\rangle s\rho_{\mathbf{m}',\mathbf{n}'} \\
&\displaystyle -  \langle\mathbf{m}|X_{+,d}^{n-r}|\mathbf{m}'\rangle\langle\mathbf{n}'|X_-^{n-r}|\mathbf{n}\rangle \rho_{\mathbf{m}',\mathbf{n}'}s \bigg], 
\end{array}
\end{equation}    
which take the explicit form of the HEOM. The reduced density matrix of the system $S$ corresponds to $\rho_S = \rho_{\mathbf{0},\mathbf{0}}$. Note that, in the case of linear system-bath interaction $n=1$, the equations in Eq.~(\ref{eq:heom}) reduce to the free-pole HEOM, cf., Eq.~(5) in the supplemental material of Ref.~\cite{FreePoles}. 

\section{Construction of the purified pseudomode model for simulating the fluorescence spectrum of quantum dots}\label{sec:example2}
Here we summarize the procedure for the construction of the purified pseudomode model for the quantum dot system described in section \ref{sec:dot}. The procedure consists of the following two steps.

\begin{enumerate}
\item[Step 1.] Generate the spectral decomposition in Eq.(\ref{eq:Cdecomppos}) using the AAA algorithms. Other available algorithms can be found in Ref.~\cite{10.1063/5.0209348}. This will produce the parameters $w_l$, $\nu_l$ and $\gamma_l$ required for the next step.
\item[Step 2.] Substitute the obtain parameters $w_l$, $\nu_l$ and $\gamma_l$ into Eq.(\ref{eq:ppmeom}) and Eq.~(\ref{eq:ppmLn}) to define the purified pseudomode model. 
\end{enumerate}
Once the model is build, one can simulate its dynamics to calculate physical quantities of interest under the initial condition $\rho_\text{ppm}(0)=\rho_S\otimes\rho_\text{vac}$ with $\rho_\text{vac}=\otimes_\alpha|0_\alpha\rangle\langle0_\alpha|$ the vacuum state of all pseudomodes.

\bibliography{refs}

\end{document}